\documentstyle[12pt]{article}
\def\Eq#1{{\begin{equation} #1 \end{equation}}}
\scrollmode

\begin{document}
\footskip 1.0cm
\thispagestyle{empty}
\begin{flushright}
ISU-NP-94-12\\
UAHEP949\\
October 1994\\
\end{flushright}
\begin{center}{\LARGE 4D EDGE CURRENTS\\
         FROM\\
      5D CHERN-SIMONS THEORY\\}
\vspace*{10mm}
{\large   K. S. Gupta $^{1}$ and
          A. Stern $^{2}$ \\ }
\newcommand{\bc}{\begin{center}}
\newcommand{\ec}{\end{center}}
\vspace*{10mm}
 1){\it Department of Physics and Astronomy, Iowa State University,\\
   Ames, IA 50011, USA,}\\
\vspace*{5mm}
 2){\it Department of Physics and Astronomy, University of Alabama, \\
Tuscaloosa, Al 35487, USA.}\ec

\vspace*{10mm}
\normalsize
\centerline{\bf ABSTRACT}
\vspace*{7mm}
A class of two dimensional conformal field theories is known to
correspond to three dimensional Chern-Simons theory.  Here we claim
that there is an analogous class of
four dimensional field theories corresponding to
five dimensional Chern-Simons theory.
The four dimensional theories give a coupling between a scalar field
and an external divergenceless vector field and they may have some
application in magnetohydrodynamics.  Like in conformal
 theories they possess a
diffeomorphism symmetry, which for us is
 along the direction of the vector field,
and their generators are analogous to Virasoro generators.
Our analysis of the abelian Chern-Simons system
uses elementary canonical methods for the
quantization of field theories defined on manifolds with boundaries.
Edge states appear for these systems and they yield
a four dimensional current algebra.
We examine the quantization of these algebras in several special cases
and claim that a renormalization of the $5D$ Chern-Simons coupling
is necessary for removing divergences.

\newpage
\setcounter{page}{1}
 \baselineskip=24pt

\section{Introduction}

The connection between three dimensional
topological field theory
and two dimensional conformal field theory is well known.\cite{wit}
The latter is derivable from the former when the three dimensional domain
for the topological theory has a boundary.  Associated with this
boundary are the so-called ``edge states" which carry
representations of a Kac-Moody algebra\cite{go}.
They have been shown to be relevant for the
quantum Hall effect. \cite{zhang,blok,Bal}

Analogous edge states can also appear in higher dimensional topological
field theory but not so much is known about these systems.  Dynamics for
higher dimensional topological field theory may
be given by either a BF Lagrangian or a Chern-Simons Lagrangian.
Edge states resulting from
BF theory defined on a four dimensional space-time
manifold with a boundary
have been examined recently by members of the Syracuse group.\cite{Bal2}
The relevance of these edge currents to the scaling limit  of a
superconductor was noted.

Here we analyze Chern-Simons theory defined on a five
dimensional manifold with a boundary.  After using canonical
techniques developed in previous work along with Balachandran
and Bimonte\cite{Bal} and imposing suitable boundary conditions,
edge states are seen to appear on the four dimensional space-time
boundary.  We shall argue that these edge states may be of some relevance
in plasma physics.

Five dimensional Chern-Simons theory has been studied by a number
of authors.\cite{fpr,ns,tr}
  In particular, Floreanini, Percacci and Rajaraman in
ref. \cite{fpr} have examined Chern-Simons theories
 on five-manifolds $M=D\times R^1$,
 $D$ being a four dimensional disc and $R^1$ accounting for time.
 More specifically $ D$ was taken to be a four dimensional solid ball
 and consequently $\partial D$ was $ S^3 $.
 They then obtained a current algebra for the resulting
edge states which was expressible in terms of field strengths
$F_{ij}$ evaluated at the boundary.  A question remaining is what is
the role of these fields strengths.  If they are to be considered as
dynamical quantities we need to know their Poisson brackets (or more
precisely, their Dirac brackets as second class constraints are
present in the system.)

In section 2 we shall reproduce the current algebra of ref \cite{fpr}.
We find the added condition that the field strengths at $\partial D$
should be considered as ${\it nondynamical}$ or external fields.
The condition results from demanding that the observable currents
are everywhere differentiable.  (Dropping such a requirement
would lead to inconsistencies in the evaluation of the Poisson
brackets.)  Furthermore our results are general
in that they apply to arbitrary four-discs $D$.

Because $F_{ij}$ is fixed at $\partial D$, only one scalar
degree of freedom survives at the boundary and it corresponds to $U(1)$
gauge transformations.  We have found an effective four-dimensional
field theory description of this system and it is analogous to
a two dimensional conformal field theory.  The effective Lagrangian
yields the coupling of a scalar field $\phi(x)$
with the external vector field
$B_k=\epsilon_{ijk}\;F_{ij}|_{\partial D}$.
Since \Eq{ \vec \bigtriangledown \cdot \vec B=0\;,\label{divB}}
$B_i$ might describe a
magnetic field or the velocity vector field of an incompressible
fluid.

The equation of motion following from the effective action is
\Eq{\vec B \cdot \vec \bigtriangledown \partial_0 \phi=0\;,\label{eomea}}
$ \partial_0 $ being a time derivative.  It states that
$ \partial_0 \phi$ is constant along $B-$field lines.
Such a situation may be encountered in magnetohydrodynamics.
One possible example is that of an incompressible fluid constrained to
flow in the same direction as an external
magnetic field $\vec B(\vec x)$.
If we then set the velocity vector field
$\vec v$ of the fluid equal to
 $\partial_0 \phi \vec B$, then the equation of motion (\ref{eomea})
follows from (\ref{divB}) and the condition of
incompressibility $ \vec \bigtriangledown \cdot \vec v=0\;.$

Another example of a system described by (\ref{divB}) and
(\ref{eomea}) occurs in the
theory of rotating stars.\cite{Tas}  Here one starts with a noniuniformly
rotating conducting fluid in the presence of a static magnetic field
$\vec B(x)$.  The field is assumed to be `poloidal' which means
that in cylindrical coordinates $(r,\theta, z)$ it has the form
\Eq{ \vec B(x) = (B_r(r,z), 0 , B_z(r,z)\;)\;,\label{Box}}
subject to (\ref{divB}).
The $z-$axis corresponds to the axis of rotation.  The angular
velocity $\omega(x)$ of the fluid is known to be governed by
the isorotation law (due to Ferraro\cite{fer}), which is just Eq.
(\ref{eomea}) with the identification that $\partial_0\phi =\omega\;.$
This result
is easy to show starting from the magnetostatic Maxwell equations
and the relation for the electric current
$\vec J$ in terms of the electromagnetic fields
\Eq{\vec J=\sigma(\vec E + \vec v \times \vec B) \;, \label{JEB}}
$\sigma  $ being the conductivity.
By taking the curl of (\ref{JEB}) and using
$ \vec \nabla \times \vec E =0 $, one gets
\Eq{\vec \nabla \times \vec J= \frac{1}{\sigma} \vec \nabla
\sigma \times \vec J +\sigma \vec \nabla \times (
 \vec v \times \vec B) \;,\label{curlJ} }
{}From the form of $\vec B(x)$ in (\ref{Box}) and using
$ \vec \nabla \times \vec B =4\pi \vec J $, only the last term in
(\ref{curlJ}) contributes in the polar (or $\theta-$) direction.  Thus
\Eq{\biggl(\vec \nabla \times (
 \vec v \times \vec B)\biggr)_\theta =0\;. }
This equation is equivalent to (\ref{eomea}) upon setting the
$\theta-$component of $\vec v$ equal to $  r \partial_0 \phi $
and using (\ref{divB}).

Since the equation of motion (\ref{eomea}) states that
$ \partial_0 \phi$ is constant along $B-$field lines it follows that
the solutions are unaffected by diffeomorphisms along the $B$
direction.  Furthermore, our effective action is invariant under such
a restricted set of diffeomorphisms.
These transformations
 are the analogue of two dimensional conformal transformations.
We have found the canonical expression for the generators of these
diffeomorphisms in terms of the Chern-Simons fields.

 In this article we also take up
  the quantization for five dimensional Chern-Simons theory in some
  simple cases (although the physical meaning of such an activity
is not clear with regard to the above mentioned plasma physics
 applications).  The resulting quantum mechanical commutation relations
 for the diffeomorphism
  generators are analogous to the Virasoro algebra.

To perform the quantization of this system we must specify two things:
{\it i)} the topology of the boundary $\partial D$ and
 {\it ii)} the external magnetic field $\vec B(\vec x)$.
After specifying {\it i)} we can
write down an appropriate Fourier decomposition.  This is done in
Section $3$.  Here we shall look at two cases:
A) $\partial D$ equal to the $3-$torus $T^3$ (
and hence $D$ equal to a solid $3-$torus $T^3$)
and B)  $\partial D=S^2 \times S^1$.

With regards to {\it ii)} we initially let
$\vec B(\vec x)$ be arbitrary for case A)
and obtain the classical current algebra
and diffeomorphism algebra.  For the latter
 a standard Sugawara construction
can be employed to obtain the diffeomorphism generators.
For case B) we take
$\vec B(\vec x)$ to be a magnetic monopole field.
Our formalism applies for this system and allows for a quantization
despite the use of an earlier assumption that two form
constructed from $F_{ij}$ is exact.
We obtain the classical current algebra
and diffeomorphism algebra also for this case.
Here the analogue of the Sugawara construction
for the diffeomorphism generators involves the $3-j$ symbols.
We find that no quantization of the magnetic charge is required for a
consistent quantization of this system.

The quantization is carried out in Section 4.  Here we first look at
the case of constant functions $\vec B(\vec x)$ on $T^3$.  We obtain a
standard Fock space representation for the system.  As in conformal
field theories, the quantum diffeomorphism algebra is seen have a central
term.  However here we find that the central term is divergent due
to a sum over an infinite number of central charges.  The result
is analogous
to defining the Virasoro algebra for strings in an infinite dimensional
space-time.  The divergence in the central term
can be absorbed away by renormalizing
the Chern-Simons coupling-but at the expense of loosing the noncentral
term in the diffeomorphism algebra.  The resulting algebra is thus
 a trivial one.  We speculate that the same
conclusion applies for any choice of {\it i)} and {\it ii)}.

In sec. 5 we give some concluding remarks concerning
 possible generalizations of this work.

\section{The Canonical Formalism}

We begin with the five-dimensional Chern-Simons action which is given by
\Eq{S=\kappa\int_M A\wedge F \wedge F \label{tcsa}   }
where $F={1\over 2} F_{\mu\nu} dx^\mu dx^\nu=dA$ is the field strength
two form.  We choose the manifold $M$ to be $D\times R^1$,
with $D$ being a four dimensional disc and $R^1$ corresponding to time.
The Lagrangian density is
\Eq{{\cal L}={\kappa\over 4} \epsilon^{\mu\nu\lambda\rho\sigma}
A_\mu F_{\nu\lambda} F_{\rho \sigma }  \;.   }
It leads to both primary and secondary
constraints in the Hamiltonian formalism.  The former are
\begin{eqnarray}
c^0&=&\pi^0\approx 0 \;, \label{pcc0} \\
c^i&=&\pi^i-\kappa \epsilon^{ijkl}  F_{jk} A_l \approx 0   \;,
\label{pcci}
\end{eqnarray}
where $\pi^\mu$ are the canonical momenta,
$i,j,..=1-4$ denote spatial indices and
$\epsilon^{ijkl}= \epsilon^{0ijkl}$.  Up to terms involving the
constraints the Hamiltonian is given by
\Eq{ H=-3\kappa \int_D    A_0  F \wedge F      \;,}
where we have set $A_0$ equal to zero on the boundary
${\partial D}$ of $D$ in order to eliminate the surface term.
We can now compute the secondary constraints,
\begin{eqnarray}
g^0&=&\{c^0, H\}={{3\kappa}\over 4} \epsilon^{ijkl} F_{ij}F_{kl}
\approx 0 \;, \label{scg0} \\
g^i&=&\{c^i,H\} =3\kappa \epsilon^{ijkl}  \partial_j A_0 F_{kl}
\approx 0    \;, \label{scgi}
\end{eqnarray}
(\ref{scg0}) is the analog of the Gauss law constraint in three
dimensional Chern-Simons theory.  From it we get that
$H$ is `weakly' zero and that there are no further secondary constraints.

We next take up the constraint analysis.

\subsection{ Constraint Analysis}

We can consistently set $A_0=\pi^0=0$ in the above formalism,
thereby eliminating the constraints (\ref{pcc0}) and
(\ref{scgi}).  The remaining eight degrees of
freedom $A_i$ and $\pi^i$ are then subject to five constraints.
If $D$ had no boundary the latter constraints could be expressed by
(\ref{pcci}) and (\ref{scg0}).  For $D$ with a boundary,
more care must be taken in defining the five remaining
constraints.  In particular, in order to compute their Poisson brackets
we must insure that these constraints are differentiable.

To proceed we follow ref. \cite{fpr} and introduce smearing functions
$\Lambda=\Lambda(x)$ and $\Sigma=\Sigma(x)$ on $D$, where $\Lambda$ is a
scalar function and $\Sigma$ is a one form, $\Sigma=\Sigma_i dx^i$.
The constraints (\ref{pcci}) and (\ref{scg0}) are then replaced by
\begin{eqnarray}
G(\Lambda)=\int_D d^4x\; \Lambda g^0 &=&
3\kappa \int_D \Lambda F\wedge F   \approx 0  \;,\\
C(\Sigma)= \int_D d^4x\; \Sigma_i c^i&=&
\int_D d^4x\; \Sigma_i \pi^i
- 2\kappa \int_D \Sigma \wedge A \wedge F \approx 0     \;.
\end{eqnarray}
The requirement that $G(\Lambda)$ and $ C(\Sigma)$
be differentiable means that variations in $A_i$ and $\pi^i$
produce no surface terms.  This is obviously the case for
variations in $\pi^i$.
  By varying $A$ in $G(\Lambda)$ and $C(\Sigma)$ we find
\begin{eqnarray}
\delta G(\Lambda)&=&-6\kappa \int_D d\Lambda \wedge F\wedge \delta A
+6\kappa \int_{\partial D}  \Lambda F\wedge \delta A \;, \label{delG}  \\
\delta C(\Sigma)&=&
 2\kappa \int_D (d\Sigma \wedge A - 2\Sigma \wedge F)\wedge \delta A
-2\kappa \int_{\partial D}\Sigma \wedge A\wedge \delta A \;.\label{delC}
\end{eqnarray}
One can now classify all possible boundary conditions
for the fields and distributions consistent with the requirement
that the surface terms in (\ref{delG}) and (\ref{delC}) vanish.
We find however that only one set of such boundary conditions leads to
a nontrivial current algebra.  It is the following:
 For the distributions $\Lambda$ and $\Sigma$ we take
\Eq{ \Lambda |_{\partial D} = \Sigma |_{\partial D} =0 \;.\label{bcd}}

As in three dimensional Chern-Simons theory\cite{Bal},
 imposing boundary conditions on the distribution functions
is sufficient for the differentiability of the constraints.
However, in order to have
differentiable observables with a nontrivial current algebra
we find it necessary to also impose boundary conditions
on $A_i$.  Imposing that the potentials vanish on the boundary is too
strong as it then leads to trivial results.  A weaker condition, which
is the one we shall assume,
is the requirement that the only dynamical degrees of freedom in
 $ A$ on the boundary
${\partial D}  $ correspond to $U(1)$ gauge degrees of freedom,
or more generally shifts in $A$ by a closed form on ${\partial D}  $ ,
\Eq{ A|_{\partial D} \rightarrow
A|_{\partial D} + \omega \quad{\rm where}\quad d\omega =0 \;.\label{bcf}}
(\ref{bcf}) implies that the $U(1)$ curvature is fixed on the boundary.

We next obtain the algebra of the constraints.
With the boundary terms in (\ref{delG}) and (\ref{delC}) vanishing, we
get the following variational derivatives of
$G(\Lambda)$ and $ C(\Sigma)$:
\begin{eqnarray}
{{\delta G(\Lambda)}\over {\delta A_l}} &=&
-3\kappa \epsilon^{ijkl} \partial_i \Lambda F_{jk}   \;,
\qquad \qquad \quad  \;
{{\delta G(\Lambda)}\over {\delta \pi^i}} = 0   \;,   \\
{{\delta  C(\Sigma)}\over {\delta A_l}} &=&
-2\kappa \epsilon^{ijkl} (\Sigma_i F_{jk}- \partial_i \Sigma_j A_k)  \;,
\quad
{{\delta  C(\Sigma)}\over {\delta \pi^i}} = \Sigma_i   \;.
\end{eqnarray}
Using these derivatives we obtain the following algebra:
\begin{eqnarray}
\{G(\Lambda),G(\Lambda')\}&=&0        \;,\label{ca1} \\
\{G(\Lambda), C(\Sigma)\}&=&-6\kappa \int_D d\Lambda \wedge
\Sigma \wedge F             \;,\\
\{ C(\Sigma), C(\Sigma')\}&=&-6\kappa \int_D
\Sigma \wedge\Sigma' \wedge F  \;,   \label{ca3}
\end{eqnarray}
where we have assumed that all the smearing functions $\Lambda,\;
\Sigma,\;\Lambda'$ and $\Sigma'$ vanish at $\partial D$.

{}From (\ref{ca1}-\ref{ca3}) the
following linear combination of constraints is first class:
\begin{eqnarray}
 {\cal G} (\tilde\Lambda)  &=& C(d\tilde\Lambda) - G(\tilde\Lambda)\cr
&=&\int_D d^4x\; \partial_i \tilde\Lambda \pi^i -
\kappa \int_D \tilde\Lambda F\wedge F \approx 0 \;.\label{fcu1}
\end{eqnarray}
Since $\tilde\Lambda$ in (\ref{fcu1})
appears as a smearing function for $G$, and since $d\tilde\Lambda$
appears as a smearing function for $C$, we need that
\Eq{ \tilde \Lambda |_{\partial D} =d \tilde \Lambda |_{\partial D} =0
\;.}
The first class constraint (\ref{fcu1})
induces $U(1)$ gauge transformations which vanish on $\partial D$:
\begin{eqnarray}
\delta_{\tilde\Lambda} A_i&=&
\{A_i,{\cal G}(\tilde\Lambda)\}=\partial_i\tilde\Lambda\;,\cr
\delta_{\tilde\Lambda} \pi^i& =&\{\pi^i, {\cal G}(\tilde\Lambda)\} =
\kappa \epsilon^{ijkl}\partial_j\tilde\Lambda F_{kl}   \;,\label{gt2}
\end{eqnarray}
In the subsection which follows we obtain generators of $U(1)$
gauge transformations which are nontrivial on $\partial D$.
These generators are in fact the observables for this system.

In addition to (\ref{fcu1}) are the first class constraints:
\Eq{ {\cal C}(\tilde\Sigma)=C( \tilde\Sigma_i) \approx 0
\;,\label{fcsh}}
 where here the distribution $\tilde\Sigma$ is not independent
of the fields, but rather satisfies
\Eq{\tilde\Sigma\wedge F=0\label{Sigf}}
 everywhere on $D$ and $\tilde\Sigma$ vanishes
on $\partial D$.  Thus for such distributions we have
${\cal C}(\tilde\Sigma)=  C(\tilde\Sigma)  \;.$
Furthermore ${\cal C}(\tilde\Sigma)$ will have nontrivial
variations with respect to $A_i$ as well as $\pi^i$,
\Eq{  {{\delta {\cal C}(\tilde\Sigma)}\over {\delta A_l}} =
2\kappa \epsilon^{ijkl}  \partial_i \tilde \Sigma_j A_k\;,  \quad
{{\delta  {\cal C}(\tilde\Sigma)}\over {\delta \pi^i}} =\tilde
 \Sigma_i   \;.  }
The first class constraints (\ref{fcsh})
induce the following transformations on $A_i$ and $\pi^i$:
\begin{eqnarray}
\delta_{\tilde\Sigma} A_i&=&
\{A_i, {\cal C}(\tilde\Sigma)\} =\tilde\Sigma_i \;,\cr
\delta_{\tilde\Sigma} \pi^i&=&\{\pi^i, {\cal C}(\tilde\Sigma)\} =
2\kappa \epsilon^{ijkl}  \partial_j \tilde \Sigma_k A_l  \;. \label{tra2}
\end{eqnarray}
They too vanish on $\partial D$.

As noted in \cite{fpr}, the field strength matrix $[ F_{ij}] $
 satisfying the constraint $g^0=0$ is degenerate with a maximum
rank of two.  The same is true for the dual of $ F_{ij}$.
Therefore there are at least two independent solutions
 $\tilde\Sigma=\tilde\Sigma^{(1)},\tilde\Sigma^{(2)}$ to (\ref{Sigf}),
 and there are at least two first class constraints
of the form (\ref{fcsh}).  A linear combination of
these constraints along with the first class constraint
(\ref{fcu1}) was shown to generate
diffeomorphisms in the interior of $D$.  It is given by
\Eq{ \Delta(w^{(0)})={\cal C}(i_{ w^{(0)} } F)+
{\cal G}(i_{ w^{(0)}} A)\;, \label{difgen}  }
where $w^{(0)}$ is a vector valued function which is
required to vanish on $\partial D$ and $i_{ w^{(0)}}$ denotes contraction
with $w^{(0)}$.  The distribution $\tilde\Sigma =i_{ w^{(0)}}F$
satisfies (\ref{Sigf}) due to
\Eq{\epsilon^{ijk\ell} (i_{w^{(0)}}F\wedge F)_{jk\ell}
=3\epsilon^{ijk\ell} F_{mj}F_{k\ell}w^{(0)m}
={3\over 4}\epsilon^{mjk\ell} F_{mj}F_{k\ell}w^{(0)i} \;,\label{FwF}}
which is (weakly) zero.

\subsection{Observables and Current Algebra}

In the above, the phase space is spanned by eight variables
$A_i$ and $\pi^i$ subject to five constraints.
Since at least three of the five constraints are first class,
none of the eight phase space variables survive as gauge invariant
 field degrees of freedom (or classical observables) in the
interior of $D$.  If $D$ had no boundary the number
of gauge invariant degrees of freedom for this system would be finite.

Are there any additional classical observables for
$D$ with a boundary?  Here we are asking for quantities with
the following properties.  i) The quantities should be invariant
with respect to transformations (\ref{gt2}) and (\ref{tra2}).
Equivalently, they should have zero Poisson
brackets with all first class constraints.  ii) They should be
differentiable with respect to all of the phase space variables.
iii) Finally they should be trivial
 for the case of no boundary $\partial D$.

One set of quantities satisfying all of the above properties is:
\Eq{q_1(f) = \int_D d^4x\; \partial_i f \pi^i  \;,\label{obq1} }
where $f=f(x)$ is a scalar distribution function on $D$.
It is easy to check that
$q_1(f)$ is invariant under $U(1)$ gauge transformations (\ref{gt2}).
Furthermore, $q_1(f)$ has zero Poisson brackets with (\ref{fcsh}) and
hence property i) is satisfied.  With regards to ii)
$q_1(f)$ is clearly differentiable with respect to $A_i$ and $\pi^i$.
For this to be true
no restrictions need to be imposed on $f$ at the boundary $\partial D$,
unlike the distribution functions $\Lambda$ and $\Sigma$.  Finally
iii), if $D$ has no boundary we can write $q_1(f)$ as a linear
combination of constraints, and hence it is (weakly) zero.  Furthermore
we note that for $D$ with a boundary, any two of observables
$q_1(f)$ and $q_1(f')$ are equivalent if the distribution functions $f$
and $f'$ have the same values on $\partial D$.  This is since
 \Eq{q_1(f) - q_1(f')=C(df-df')-{2\over 3}G(f-f')\approx 0 \;,\quad
f|_{\partial D} =  f'|_{\partial D} \;.}
Thus the set of all $q_1(f)$'s with functions
$f$ coinciding on the boundary
defines an equivalence class of observables.

Another quantity with properties i-iii) is
\Eq{q_2(f) =  \kappa \int_D( fF + 2 df \wedge A)\wedge F \;,\label{q2}}
where again we assume no restrictions on distribution function
$f$ at the boundary $\partial D$.  With regards to i),
$q_2(f)$ has zero Poisson
brackets with all first class constraints.
For ii) to be true
 we need to check variations with respect to $A_i$.  We get
\Eq{\delta q_2 (f) = 2\kappa \int_D df \wedge F\wedge \delta A
+2\kappa \int_{\partial D} f d\delta A \wedge A    \;. \label{varq2}  }
The boundary term in (\ref{varq2})
vanishes upon assuming that the variations of $A$ are of the form
(\ref{bcf}).  Hence $q_2(f)$ is differentiable, and this is the
justification for the boundary condition (\ref{bcf}).  For the case of
 $D$ with no boundary we can integrate (\ref{q2}) by parts to
find $q_2(f)=-{1\over 3}G(f)$
which is (weakly) zero and thus iii) is satisfied.  As before,
$q_2(f)$ and $q_2(f')$ are equivalent if $f$
and $f'$ have the same values on $\partial D$, ie.
$q_2(f)\approx q_2(f')$ for $f|_{\partial D} =  f'|_{\partial D} \;.$

The next step is to compute the algebra of observables.
Our approach is to look for Dirac variables.  Thus we want quantities
that in addition to satisfying i-iii), have the property iv) that
their Poisson brackets with the second class constraints are zero.
Therefore they will have zero brackets with all constraints.
Their current algebra can then be computed directly
from their Poisson brackets.

We have found only one set of variables which satisfy i-iv).  They
are:
\Eq{  q(f) = q_1(f)+q_2(f)\;.    }
Thus
\Eq{ \{ G(\Lambda), q(f)\} =  \{ C(\Sigma), q(f)\} = 0 \;,}
for all $\Lambda$ and $\Sigma$.  In order to compute
 the Poisson bracket algebra of $q(f)$ we first need the
variational derivatives,
\Eq{ {{\delta q(f)}\over {\delta A_\ell}} =
 \kappa \epsilon^{ijk\ell} \partial_i f F_{jk}   \;,  \qquad
{{\delta q(f)}\over {\delta \pi^i}} = \partial_i f   \;. \label{vdqf} }
{}From them it follows that
the $q(f)$'s are generators of $U(1)$ gauge transformations
which, unlike those generated by ${\cal G} (\tilde\Lambda)$,
are nontrivial on $\partial D$.
[For distributions $f$ which vanish on $\partial D$,
$q(f)$ is identical to the first class constraint
${\cal G}(f)$.]  From (\ref{bcf})
the degrees of freedom generated by $q(f)$ coincide
with the degrees of freedom
in $A|_{\partial D}$.  The former are $U(1)$ gauge transformations
connected to the identity.

Now from (\ref{vdqf}) we get the following current algebra
 \begin{eqnarray}
 \{q(f),q(f')\}&=& 4\kappa \int_{ D} df \wedge df' \wedge F   \cr
&=&4\kappa \int_{\partial D} (fdf'-f'df) \wedge F \;.  \label{ca}
\end{eqnarray}
We have thus recovered the same algebra as in \cite{fpr}.  The two form
$F$ appearing in (\ref{ca}) has no dynamics, since as stated earlier
the $U(1)$ curvature is fixed on the boundary due to (\ref{bcf}).

\subsection{Effective Lagrangian}

Since due to the boundary condition (\ref{bcf}),
 $A|_{\partial D}$ contains only a scalar degree of freedom
it makes sense that the above Chern-Simons system has an effective scalar
field theory description.  We shall now examine such a description.

The current algebra (\ref{ca}) which was gotten starting from the
five dimensional Chern-Simons action
is also obtainable from a Lagrangian
written solely on the boundary $\partial D$.  It is
\Eq{ L = 4\kappa \int_{\partial D}\partial_0 \phi\;d \phi \wedge F\;,
\label{effl}} where $\phi$ is a scalar field.  As before
the curvature 2-form $F$ is nondynamical on $\partial D$.
In order to recover (\ref{ca}) one also needs to use that $F$ is closed.
(\ref{effl}) then represents the coupling of a scalar field
to a static divergenceless vector field $B_i=\epsilon_{ijk}F_{jk}$.
Physically $B_i$ might describe a
magnetic field or the velocity vector field of an incompressible
fluid.  The resulting equation of motion reads
 \Eq{ \partial_0 \vec \bigtriangledown \cdot ( \phi \vec B)=0\;,  }
or equivalently (\ref{eomea}).
 General solutions for $\phi$ are of the form
\Eq{ \phi(\vec x,t) =\phi_0(\vec x) +\phi_1(\vec x, t)  \quad
{\rm where} \quad \vec B \cdot \bigtriangledown \phi_1 =0\;.}
$\phi_0$ and $\phi_1$ are analogous to
the chiral modes of two dimensional conformal field theory.

We now show how to recover the current algebra (\ref{ca}) from
the Lagrangian (\ref{effl}).  In the Hamiltonian formalism,
the momenta $\Pi(x)$ conjugate to $\phi(x)$ are constrained by
\Eq{\Pi -2\kappa    \vec B \cdot \vec \bigtriangledown
\phi  \approx 0\;, \label{efc}}
and there are no further constraints.  Let us assume that $B_i$ is
nonvanishing everywhere.  Then (\ref{efc}) are second class and
they are eliminated by finding Dirac variables.
It is easy to obtain such variables which we denote by $\rho(x)$,
\Eq{\rho=\Pi +2\kappa    \vec B \cdot \vec \bigtriangledown \phi\;. }
Finally to recover algebra (\ref{ca}) one can now define
$q(f)$ according to \Eq{q(f)=\int_{\partial D} f(x) \rho(x)\;,}
and compute its Poisson brackets.

\subsection{Diffeomorphisms}

Finally we take up the topic of diffeomorphism invariance in the
five dimensional Chern-Simons theory or equivalently the four
dimensional effective field theory.  This invariance is the analogue
of two dimensional conformal symmetry.

We begin with the effective Lagrangian (\ref{effl}).
Since the curvature 2-form $F$ is nondynamical on $\partial D$,
(\ref{effl}) is not invariant under all possible
diffeomorphisms on $\partial D$, \Eq{\delta \phi = {\cal L}_W \phi\;, }
 $ {\cal L}_W $ denoting a Lie derivative.  Rather, it is only invariant
for those diffeomorphisms on $\partial D$ which are
along one direction - the $B_i$ direction,
\Eq{W_i=\epsilon(x) B_i\;,\label{bcw}}
$\epsilon(x)$ being an arbitrary function on $\partial D$.

These diffeomorphisms on $\partial D$ can be extended to all of $D$
and their generators can be expressed in terms of $A$ and $F$.
The generators are given by
\Eq{ \Delta(w)= \int_D d^4x\;({\cal L}_wA)_i c^i
- \int_D d^4x (i_w A)g^0 + 2\kappa\int_{\partial D}
(i_wA) F\wedge A \;, \label{dgen2}  }
where we require
\Eq{w_i|_{\partial D}=W_i=\epsilon(x) B_i\;.\label{bcw2}}
As in the above, $B_i=\epsilon_{ijk}F_{jk} $ is a tangent vector in
$\partial D$.  From (\ref{bcw2}) it also follows that
\Eq{i_wF|_{\partial D}=0\;.\label{iwf}}

Upon examining the expression (\ref{dgen2}) for $\Delta(w)$,
we note that we cannot write the first two terms
as linear combinations of the constraints $C(\Sigma)$ and $G(\Lambda)$
because the relevant distribution functions $\Sigma={\cal L}_w A$
 and $\Lambda=i_w A$ do not vanish on $\partial D$.
Using ${\cal L}_w=di_w+i_w d$ we can however rewrite (\ref{dgen2})
in terms of the first class constraints
${\cal C}(\tilde\Sigma)$, with $\tilde\Sigma=i_w F$,
\Eq{ \Delta(w)={\cal C}(i_w F)+\int_D d^4x\; \partial_i (i_w A) \pi^i-
\kappa \int_D(i_w A)F\wedge F \;. \label{dgen3}  }
As required for distributions of ${\cal C}(\tilde\Sigma) \;,
\tilde\Sigma=i_w F$ satisfies (\ref{Sigf}) [due to
(\ref{FwF})] and it vanishes on $\partial D$ [due to (\ref{bcw2})].
When $w|_{\partial D}=0$, the second and third term in (\ref{dgen3})
can be identified with the first class constraint ${\cal G}(i_wA)$
(since then $i_wA|_{\partial D}=0$) and we thus recover the generators
(\ref{difgen}) of difeomorphisms in the interior of $D$.
The boundary term in (\ref{dgen2}) is needed for differentiability.
Its variation in $A$ is
\Eq{ 2\kappa\;\delta\int_{\partial D}(i_wA) F\wedge A=
4\kappa\int_{\partial D}(i_wA) F\wedge\delta A \;. \label{vabt}  }
The remaining terms in (\ref{dgen2}) produce the following boundary
terms upon varying $A$:
\begin{eqnarray}
& &-2\kappa\int_{\partial D}\biggl({\cal L}_wA\wedge A +3(i_wA) F\biggr)
\wedge \delta A\cr &=& -2\kappa\int_{\partial D}
\biggl(di_wA\wedge A +i_wF \wedge A +3(i_wA) F\biggr)  \wedge \delta A\cr
&=&-2\kappa\int_{\partial D}\biggl(i_wF \wedge A +2(i_wA) F\biggr)
\wedge \delta A \;,
\end{eqnarray}
where we have used the boundary condition (\ref{bcf}) on $A$.
These boundary terms are cancelled by (\ref{vabt}) after using
the boundary condition (\ref{iwf}).

We can now compute the variational derivatives of $\Delta(w)$.  We find:
\begin{eqnarray}
 {{\delta \Delta(w)}\over {\delta A_\ell}}&=&
 \kappa \epsilon^{ijk\ell} \biggl( 2\partial_i(w^mF_{mj})A_k +
\partial_i(i_wA)F_{jk} \biggr) - {1\over 3}w^\ell g^0 +
\partial_i w^\ell c^i - \partial_i(w^i c^\ell) \;,\cr
{{\delta \Delta(w)}\over {\delta \pi^i}}&=&({\cal L}_wA)_i
 \;. \label{vdDwp}
\end{eqnarray}
We can use the variational derivatives to obtain the following Poisson
brackets of $\Delta(w)$ with the constraints:
\begin{eqnarray}
 \{\Delta(w),G(\Lambda)\}&=& G(i_wd\Lambda) \;,        \cr
 \{\Delta(w),C(\Sigma)\}&=& C({\cal L}_w\Sigma) \;.
\end{eqnarray}
Thus $\Delta(w)$ have (weakly) zero brackets with all the
constraints.  Vanishing Poisson brackets with the first class constraints
implies that $\Delta(w)$ are invariant with respect to transformations
(\ref{gt2}) and (\ref{tra2}), while vanishing Poisson brackets with
the second class constraints implies that $\Delta(w)$
are Dirac variables, or equivalently, their Dirac brackets are equal
to their Poisson brackets.
Now using (\ref{vdqf}) and (\ref{vdDwp}) we can compute the Poisson
brackets of $\Delta(w)$ with the observables $q(f)$.  We find
\begin{eqnarray}
 \{\Delta(w),q(f)\}&=&\int_D d^4x\biggl(\partial_i(i_wdf)c^i- (i_wdf) g^0
\biggr) + 4\kappa \int_{\partial D} (i_wA) F \wedge df \cr
& &        \cr
&=&q(i_w df) = q({\cal L}_w f) \;.  \label{delq}
\end{eqnarray}
For the Poisson
brackets of $\Delta(w)$ with $\Delta({w'})$ we find
\begin{eqnarray}
 \{\Delta(w),\Delta({w'})\}    &=&   \int_D d^4x
\biggl(({\cal L}_{w'}A)_j \bigl( \partial_i w^j c^i- \partial_i(w^ic^j)
\bigr)  - (w \rightleftharpoons {w'}) \biggr)  \cr
& & -  \int_D d^4x([i_w,{\cal L}_{w'}]A) g^0 +
{1\over 3} G(i_{w'} i_w F) \cr
& &+ 2\kappa \int_D \biggl( i_{w'}F\wedge A \wedge d(i_w F)
+ d(i_wA)\wedge F\wedge d(i_{w'} A) -
(w \rightleftharpoons {w'}) \biggr)  \cr
& &     \cr
 &=&  \int_D d^4x   \biggl( d [i_w,{\cal L}_{w'}]A
+ [i_w,{\cal L}_{w'}]F\biggr)_i\;c^i
 -\int_D d^4x([i_w,{\cal L}_{w'}]A) g^0\cr
& & +2\kappa \int_{\partial D}([i_w,{\cal L}_{w'}]A) F\wedge A    \cr
& &  \cr
&=&  \int_D d^4x\;({\cal L}_{{\cal L}_w{w'}}A)_i c^i
- \int_D d^4x (i_{{\cal L}_w{w'}} A)g^0
 + 2\kappa\int_{\partial D} (i_{{\cal L}_w{w'}}A) F\wedge A
    \cr
& &  \cr
&=& \Delta({{\cal L}_w{w'}})    \label{deldel}
\end{eqnarray}
where ${w'}_i|_{\partial D}=\epsilon'(x) B_i$ and we utilized the
identity $i_{{\cal L}_w{w'}} A=[i_w,{\cal L}_{w'}]A$.
The Poisson brackets (\ref{delq}) and (\ref{deldel}) are
identical in form to the analogous ones in three dimensional Chern-Simons
theory \cite{Bal}.
\medskip

\section{Mode Expansion}

The remainder of this article is devoted to the quantization
of this system.

For quantization we must specify what is the boundary $\partial D$ of
$D$ and also the static two-form $F$ on $\partial D$.
As a prelude we first perform mode expansion of the current algebra
(\ref{ca}) as well as the diffeomorphism algebra (\ref{deldel}).
We shall give the analogue of the Sugawara construction
for the diffeomorphism generators.

We examine two cases:  A)
$\partial D=T^3\;(=S^1 \times S^1\times S^1)$ with $F|_{\partial D}$
initially arbitrary and  B)  $\partial D=S^2 \times S^1$ with
$F|_{\partial D}$ equal to a magnetic monopole field.
In our above canonical treatment $F$ was assumed to be exact, and this
will also be the case for our first example.  However
in the second example,  $F|_{\partial D}$ is closed but not exact.
We will show that our formalism can be consistently
extended to this case.

\subsection{ Case A) $\partial D=T^3$ and
$F|_{\partial D}$ arbitrary }

For a basis of test functions on the three-torus we choose
\Eq{\chi^{(\vec N)}(\vec \theta) =
\exp{\{i\vec N \cdot \vec \theta\}} \;,\quad
\vec N = (N_1,N_2,N_3)\;, \quad
\vec \theta = (\theta_1,\theta_2,\theta_3)\;,  }
with $N_i=$ integers and $0\le\theta_i < 2\pi$.  Then an arbitrary
$F|_{\partial D}$ can be expanded according to
\Eq{ F|_{\partial D }=\sum_{\vec N} \epsilon^{ijk} b_i^{(\vec N)}
\chi^{(\vec N)}(\vec \theta)\; d\theta_j\wedge d\theta_k \;,\label{Fexp}}
where $ \vec  b^{(\vec N)}=
(  b_1^{(\vec N)}, b_2^{(\vec N)}, b_3^{(\vec N)})$
are constant vectors.  From the closedness condition on $F$ they satisfy
\Eq {\vec N \cdot  \vec  b^{(\vec N)}= 0 \;,\label{NdotV}}
for all $\vec N$.

In considering the observables $q(f)$, let
$f^{(\vec N)}$ be a distribution function defined on all $D$ with the
boundary value given by
\Eq{f^{(\vec N)}|_{\partial D}   =\chi^{(\vec N)}(\vec \theta)\;.
\label{bcfN}}
All $q(f^{(\vec N)})$ for $f^{(\vec N)}$ satisfying the boundary
condition (\ref{bcfN}) are (weakly) equivalent.  Thus
$q_{\vec N} = q(f^{(\vec N)})$ defines an equivalence class of
observables.   Using (\ref{ca}) and (\ref{Fexp}) we can compute the
Poisson bracket algebra for the $q_{\vec N}$'s.  We find:
\Eq{\{q_{\vec N}\;,\; q_{\vec M}\}=(4\pi)^3i\kappa \;(\vec M-\vec N)\cdot
\vec b^{(-\vec N - \vec M)} \;. \label{qNqM}}
{}From (\ref{NdotV}) it then follows that $q_{\vec N=(0,0,0)}$ is a
central charge for any $\vec b^{(\vec N)}$.
Whether or not more such central charges exist depends upon
the values for $\vec b^{(\vec N)}$.

For the case of the diffeomorphism generators $\Delta(w)$, let
$\vec w^{(\vec N)}$ be a vector-valued distribution function
on $D$ with a boundary value given by
\Eq{w_i^{(\vec N)}|_{\partial D}   = {1\over 4}
\chi^{(\vec N)}(\vec \theta)\;\epsilon_{ijk} F^{jk}|_{\partial D}=
\sum_{\vec M} b_i^{(\vec M +\vec N)}
\chi^{(\vec M)}(\vec \theta)\;.
\label{bcwi}}
All $\Delta_{\vec N}=\Delta(w^{(\vec N)})$
for $w^{(\vec N)}$ satisfying the boundary
condition (\ref{bcwi}) are (weakly) equivalent and
define an equivalence class of diffeomorphism generators.
{}From (\ref{delq}) and (\ref{deldel}) their Poisson brackets with
$q_{\vec N}$ and with themselves are given by
\begin{eqnarray}
\{\Delta_{\vec N}\;,\; q_{\vec M}\}&=&\sum_{ \vec P} i\vec M \cdot
\vec b^{(\vec P)} q_{\vec M +\vec N + \vec P}  \;,  \\
\{\Delta_{\vec N}\;,\; \Delta_{\vec M}\}&=&\sum_{\vec P} i(\vec M-\vec N)
\cdot  \vec b^{(\vec P)} \Delta_{\vec M +\vec N + \vec P}  \;.
\end{eqnarray}
These relations are realized if we set $\Delta_{\vec N}$ equal to
\Eq{L_{\vec N}=
{1\over{4^4 \pi^3\kappa}}\sum_{\vec M}  q_{\vec M} q_{\vec N-\vec M}
\label{csc}}
and then apply the Poisson brackets (\ref{qNqM}).  This corresponds
to the Sugawara construction of the Virasoro generators.
As in three dimensional Chern-Simons theory\cite{Bal} we expect
that $\Delta_{\vec N}\approx L_{\vec N}$.

\subsection{ Case B) $\partial D=S^2 \times S^1$ with
$F|_{\partial D}$ equal to a magnetic monopole field}

We let $\theta$ and $\phi$ be the polar and azimuthal angles
parametrizing $S^2$ with $\psi$ parametrizing $S^1$.
In terms of these coordinates the magnetic monopole two form is given by
\Eq{F|_{\partial D} =g \sin \theta d\theta \wedge d\phi  \;,\label{MF}}
$g$ being a constant magnetic charge.
We note that it is possible to smoothly extend the two form $F$ to the
interior of the four dimensional disc
$D$ without introducing any singularities.  For this
let us introduce the radial coordinate $r$, $0\le r \le 1$, with $r=1$
defining the boundary $\partial D$.  We define a function $\hat g(r)$
such that
\Eq{ \hat g \rightarrow  0 \;,\quad {{d\hat g}\over {dr}}\rightarrow 0\;,
\quad {\rm as}\;\; r\rightarrow  0\;,}
\Eq{ \hat g \rightarrow  g \;,\quad {{d\hat g}\over {dr}}\rightarrow 0\;,
\quad {\rm as}\;\; r\rightarrow  1\;.}
Now we can take \Eq{F =\hat g(r) \sin \theta d\theta \wedge d\phi
- {{d\hat g}\over{dr}} \cos \theta dr \wedge d\phi \;, \label{FfaD} }
for all of $D$.  (\ref{FfaD})
 is a closed two form which coincides with (\ref{MF})
at $r=1$ and is nonsingular at $r=0$.

For a basis of test functions on $\partial D$ we choose
\Eq{Y^m_\ell(\theta,\phi) e^{in\psi}\;,\label{bvfdf}}
where $\ell=0,1,2...$, $m=-\ell, -\ell +1,...\ell$ and $n$ are integers.
$ Y^m_\ell(\theta,\phi) $ are the standard spherical harmonics.  We let
$f^{(n,m,\ell)}$ be a distribution function defined on all $D$ with the
boundary value given by (\ref{bvfdf}).
$q_{n,\ell,m} = q(f^{(n,\ell,m)})$ defines an equivalence class of
observables.   Using (\ref{ca}) and (\ref{MF}) we can compute the
Poisson bracket algebra for the $q_{n,\ell,m}$'s.  We find:
\Eq{\{q_{n,\ell,m}\;,\; q_{n',\ell',m'}\}=
-16\pi ig\kappa \; (-1)^m n\; \delta_{n,-n'}
  \delta_{\ell,\ell'} \delta_{m,-m'}   \;.\label{qqmp}}
Here $q_{0,\ell,m}$ are central charges and they carry
representations of the rotation group.

Concerning the diffeomorphism generators $\Delta(w)$, let
$\vec w^{(n,\ell,m)}$ be a vector-valued distribution function
on $D$ with a boundary value given by
\Eq{w_{\theta}^{(n,\ell,m)}|_{\partial D}   =
w_{\phi}^{(n,\ell,m)}|_{\partial D}   =  0\;,\quad
w_{\psi}^{(n,\ell,m)}|_{\partial D}   =
Y^m_\ell(\theta,\phi) e^{in\psi}\; \;.\label{bcwi2}}
Then $\Delta_{n,\ell,m}=\Delta(w^{(n,\ell,m)})$
for $w^{(n,\ell,m)}$ satisfying the boundary condition (\ref{bcwi2})
define an equivalence class of diffeomorphism generators.
Their Poisson brackets are given by
\begin{eqnarray}
  \{\Delta_{n,\ell,m}\;,\; q_{n',\ell',m'}\} &=& in'
\sum_{\ell''=|\ell-\ell'|}^{|\ell+\ell'|}\;\sum_{m''=-\ell''}^{\ell''}
\alpha^{\ell,\ell',\ell''}_{m,m',m''} \;  q_{n+n',\ell'',m''}   \;,
\label{Dqmp} \\
 \{\Delta_{n,\ell,m}\;,\; \Delta_{n',\ell',m'}\} &=& i(n'-n)
\sum_{\ell''=|\ell-\ell'|}^{|\ell+\ell'|}\;\sum_{m''=-\ell''}^{\ell''}
\alpha^{\ell,\ell',\ell''}_{m,m',m''} \;  \Delta_{n+n',\ell'',m''}   \;.
\label{DDmp}
\end{eqnarray}
In obtaining the above expressions we used
 the composition relation for spherical harmonics,
 \Eq{Y^{m}_{\ell}(\theta,\phi) Y^{m'}_{\ell'}(\theta,\phi)
=\sum_{\ell''=|\ell-\ell'|}^{|\ell+\ell'|}\;\sum_{m''=-\ell''}^{\ell''}
\alpha^{\ell,\ell',\ell''}_{m,m',m''} \;Y^{m''}_{\ell''}(\theta,\phi)
\;.}
The $\alpha$ coefficients are given by
\Eq{\alpha^{\ell,\ell',\ell''}_{m,m',m''} =
(-1)^{m''}  \sqrt{{{(2\ell+1)(2\ell'+1)(2\ell''+1)}\over {4\pi}}}
\pmatrix{ \ell & \ell' & \ell'' \cr 0 & 0 & 0 }
\pmatrix{ \ell & \ell' & \ell'' \cr m & m' & -m'' }    \;,}
$\pmatrix{ \ell & \ell' & \ell'' \cr m & m' & -m'' } $  being
the $3-j$ symbols.  The $\alpha$'s satisfy the symmetry properties:
\begin{eqnarray}
 \alpha^{\ell,\ell',\ell''}_{m,m',m''}&=&
    \alpha^{\ell',\ell,\ell''}_{m',m,m''} \;,\cr
& & \cr
\alpha^{\ell,\ell'',\ell'}_{m,-m'',-m'}&=& (-1)^{m'+m''}
\alpha^{\ell,\ell',\ell''}_{m,m',m''}  \;  \;,  \cr
& & \cr
\sum_{\ell',m'}  \alpha^{\ell_1,\ell_2,\ell'}_{m_1,m_2,m'}  \;
\alpha^{\ell',\ell_3,\ell}_{m',m_3,m} &=&
\sum_{\ell',m'}  \alpha^{\ell_3,\ell_2,\ell'}_{m_3,m_2,m'}      \;
\alpha^{\ell',\ell_1,\ell}_{m',m_1,m} \;.\label{ida2}
\end{eqnarray}
The last identity follows from the decomposition of three spherical
harmonics
 \Eq{Y^{m_1}_{\ell_1}(\theta,\phi) Y^{m_2}_{\ell_2}(\theta,\phi)
Y^{m_3}_{\ell_3}(\theta,\phi)=
\sum_{\ell,m,\ell',m'}  \alpha^{\ell_1,\ell_2,\ell'}_{m_1,m_2,m'} \;
\alpha^{\ell',\ell_3,\ell}_{m',m_3,m}
 \;Y^{m}_{\ell}(\theta,\phi)   \;.}
Using the identities (\ref{ida2}) along
with the Poisson brackets (\ref{qqmp}), we can verify that
(\ref{Dqmp}) and (\ref{DDmp}) are
 realized for $\Delta_{n,\ell,m}$ equal to
\Eq{L_{n,\ell,m}=
{1\over{32 \pi g\kappa}}\sum_{\ell',\ell'',n'}\;
\sum_{m'=-\ell'}^{\ell'} \; \sum_{m''=-\ell''}^{\ell''}
(-1)^{m'}\; \alpha^{\ell,\ell',\ell''}_{m,-m',m''}  \;
q_{n',\ell',m'}\; q_{n-n',\ell'',m''}  \;,\label{SugB}}
This gives the Sugawara construction for case B).

\section{Remarks on Quantization}

Here we make some remarks concerning the quantization for cases A) and
B) of the previous section.  For both cases we obtain standard Fock
space representations.
We compute the central term in the quantum diffeomorphism algebra for
one particular example and find that it is divergent.  The divergence
is due to the existence of an infinite number of central charges.
It can be absorbed away by renormalizing
the Chern-Simons coupling-but at the expense of loosing the noncentral
term in the diffeomorphism algebra.  The resulting algebra is then
 a trivial one.
Since
an infinite number of central charges exist in all of the examples
which we study
  we speculate that similar conclusions apply there as well.

We begin our discussion with case A).
For case  A) we need to know the expression for $F|_{\partial D}$.
Of course the simplest choice we could make would be to take
all of its components
(with respect to the $\theta_i$ coordinates) equal to
 constants.  This corresponds to zero windings around the torus.
 We first consider this choice.
Then we examine case A) when the components of $F|_{\partial D}$
(with respect to the $\theta_i$ coordinates) are not all equal to
 constants, thus allowing for non-zero windings around the three-torus.
As a general analysis appears to be complicated we shall assume that
only one nontrivial mode contributes to the expansion (\ref{Fexp})
for $F|_{\partial D }$.

\subsection{Case A) $-$ Zero Windings}

Constant $F_{ij}|_{\partial D }$ or zero windings means that we make
the following choice for the vectors $ \vec b^{(\vec N)} $:
\Eq{ \vec b^{(\vec N)} = \delta_{\vec N,(0,0,0)}  \vec b^{(0)} \;.
\label{V0}}
By $ \delta_{\vec N,\vec M}$ we mean
 $ \delta_{N_1,M_1} \delta_{N_2,M_2} \delta_{N_3,M_3}$.
The choice (\ref{V0}) identically satisfies (\ref{NdotV}).
{}From (\ref{qNqM}) the only nonzero
Poisson brackets for the observables are between $q_{\vec N}$ and
$q_{-\vec N}= q_{\vec N}^*$, where
$\vec N \cdot \vec b^{(0)} \ne 0 \;.$  The variables are:
\Eq{\{q_{\vec N}\;,\; q_{\vec N}^*\}= -2
(4\pi)^3i\kappa \;\vec N \cdot \vec b^{(0)} \;. \label{qNqNa}}
The observables $q_{\vec N}$, where
$\vec N \cdot \vec b^{(0)} = 0 \;,$ behave as central charges.
(For example if we set $b_2^{(0)} =b_3^{(0)}=0$, the central charges
are given by $q_{(0,N_2,N_3)}$ for all values of $N_2,\;N_3$.)

For the quantum theory we replace (\ref{qNqNa}) by
 the commutation relations
\Eq{[Q_{\vec N}\;,\; Q_{\vec N}^\dagger]=  2
(4\pi)^3 \kappa \;\vec N \cdot \vec b^{(0)} \; ,\label{QNQNa}}
where $Q_{\vec N}$ is the quantum operator associated with $q_{\vec N}$.
Thus if $ \kappa >0  \;\;( \kappa <0 ),\;Q_{\vec N}$ for
$\;\vec N \cdot \vec b^{(0)} \;>0 \;(\;\vec N \cdot \vec b^{(0)} \;
<0)$ are annihilation operators (upto a normalization) and
$Q_{\vec N}^\dagger $ creation operators.
The ``vacuum'' $| 0 >$ can be defined as usual by
\Eq{ Q_{\vec N} \mid 0> = 0 \quad\;{\rm if}~
\quad \kappa \;\vec N \cdot \vec b^{(0)} \;>0\;. }
The excitations are obtained by applying $Q_{\vec N}^\dagger$
to the vacuum.

The quantum diffeomorphism generators are normal ordered versions
of their classical counterparts (\ref{csc}), \Eq{{\bf L}_{\vec N}=
{1\over{4^4 \pi^3\kappa}}\sum_{\vec M} :Q_{\vec M} Q_{\vec N-\vec M}:\;
 \;\;.}
Let us now choose $\kappa >0$.  Then we have that ${\bf L}_{\vec N}|0>=0$
 for $\;\vec N \cdot \vec b^{(0)} \;>0 $.
{}From (\ref{V0}) the ${\bf L}_{\vec N}$'s
 yield an algebra which includes central terms,
\Eq{[{\bf L}_{\vec N},{\bf L}_{\vec M}]=
(\vec N-\vec M)\cdot \vec b^{(0)}\; {\bf L}_{\vec N+\vec M}
+ \gamma(\vec N) \;\delta_{\vec N+\vec M,(0,0,0)}  \;. }
$ \gamma(\vec N)= - \gamma(-\vec N)$ is a function of the three
dimensional lattice.  Constraints on $ \gamma(\vec N)$ come from the
Jacobi identity
$$[[{\bf L}_{\vec N},{\bf L}_{\vec M}],{\bf L}_{-\vec N-\vec M}] +
[[{\bf L}_{\vec M},{\bf L}_{-\vec N-\vec M}], {\bf L}_{\vec N}]+
[[{\bf L}_{-\vec N-\vec M}, {\bf L}_{\vec N}],{\bf L}_{\vec M}]=0\;,$$
which gives
\Eq{ (\vec N-\vec M)\cdot \vec b^{(0)}\; \gamma (\vec N+\vec M)=
(2\vec M+\vec N)\cdot \vec b^{(0)} \;\gamma (\vec N) -
(2\vec N+\vec M)\cdot \vec b^{(0)}\; \gamma (\vec M) \;\;.\label{ccga}}
One immediate consequence of this equation is that
$\gamma (\vec N) =\gamma (\vec M)$ if $\vec N\cdot \vec b^{(0)}=$
$\vec M\cdot \vec b^{(0)}\ne 0$.  Thus for
$\vec N\cdot \vec b^{(0)}\ne 0$, $\gamma (\vec N)$ can be
a function only of $\vec N\cdot \vec b^{(0)}$.

For the case of $\vec N\cdot \vec b^{(0)}=0$, the consistency
condition (\ref{ccga}) implies that
$$2\gamma(\vec N)= \gamma(\vec M) - \gamma(\vec M + \vec N)\;,\quad
{\rm for  }\;\vec M\cdot \vec b^{(0)}\ne 0\;.$$
But since $\vec M \cdot \vec b^{(0)}\ne 0$,
$\gamma (\vec M)$ is a function only of $\vec M\cdot \vec b^{(0)}$,
and furthermore it is identical to
$ \gamma(\vec M + \vec N)$.  It therefore follows that
\Eq{\gamma(\vec N)= 0\;\quad{\rm for  }\;\vec N\cdot \vec b^{(0)}= 0\;.
\label{tzm}}

Returning to the case of $n=\vec N\cdot \vec b^{(0)}\ne 0$,
(\ref{ccga}) gives a recursion relation for generating all
$\gamma (\vec N)=h(n)$, starting with the values
of $\gamma$ at different points on the three dimensional lattice.
The recursion relation for $h(n)$ is identical to that for the central
term in the Virasoro algebra defined for
the one dimensional lattice\cite{gsw}.  Its solution is well
known, \Eq{ h(n)=c_1 n + c_3 n^3 \;.\label{solh}}
[We note that for the one dimensional lattice $n$ takes
on integer values and that this is not in general true
for our case.  However, if we assume that all of the components of
$\vec b^{(0)}$ are rationally related, then our $h$ can be redefined
so that its arguments are integers and consequently
the solution (\ref{solh}) applies.  The form (\ref{solh})
for $h$ is unaffected by such a redefinition.]

It remains to calculate the coefficients $c_1$ and $c_3$.
For this it is sufficient to compute the matrix element
$<0|{\bf L}_{\vec N}{\bf L}_{-\vec N}|0>$ for two different values of
$n=\vec N\cdot \vec b^{(0)}> 0$.  Let us first choose
$\vec N= \vec N^{(1)}$ so that
$\nu_1=\vec N^{(1)}\cdot \vec b^{(0)}$ gives the smallest positive
value for $n$.  We get
$$<0|{\bf L}_{\vec N^{(1)}}{\bf L}_{-\vec N^{(1)}}|0>=2\nu_1
<0|{\bf L}_{(0,0,0)}|0>\;,$$ or \Eq{ h( \nu_1)=0\;.\label{hnu1}}
Then choose $\vec N= \vec N^{(2)}$ so that
$\nu_2=\vec N^{(2)}\cdot \vec b^{(0)}$ is the next smallest positive
value for $n$, $\nu_2>\nu_1>0$.
 Here we get two possible answers for $h(\nu_2)$
depending on whether or not the value of $\nu_2$
is twice that of $\nu_1$,
\Eq{ h(\nu_2)=\left\{ \begin{array}{ll}
{{\cal D}\over 4}  \nu_1 (\nu_2 -\nu_1) &
\mbox{for $\;\nu_2 \ne 2\nu_1$}\;,\cr   & \cr
{{\cal D}\over 2}   (\nu_1)^2 & \mbox{for $\;\nu_2 = 2\nu_1$ }\;,
\end{array} \right. \label{hnu2}}
and thus from (\ref{hnu1}) and (\ref{hnu2}),
 \Eq{ \gamma(\vec N)= h(n=\vec N\cdot \vec b^{(0)}) =
\left\{ \begin{array}{ll}
{{\cal D}\over 4} \frac{ \nu_1n [n^2 -(\nu_1)^2]}
{\nu_2 [\nu_2 +\nu_1]} &   \mbox{for $\;\nu_2 \ne 2\nu_1$}\;,\cr
 & \cr {{\cal D}\over {12} } \frac {n [n^2 -(\nu_1)^2] }{\nu_1}
  & \mbox{for $\;\nu_2 = 2\nu_1$ }\;.
\end{array} \right.\label{tct} }
We note that this answer includes the case of
$n=\vec N\cdot \vec b^{(0)}=0$ since (\ref{tzm}) is then satisfied.
 ${\cal D}$ in (\ref{hnu2}) and (\ref{tct})
 counts the number of lattice points, labeled by $\vec N^{(0)}$,
for which $\nu_0=\vec N^{(0)}\cdot \vec b^{(0)}=0$.  But since
we have an infinite lattice (and the components of
$\vec b^{(0)}$ are rationally related), ${\cal D}$ {\it is divergent}.
For this result we only need to find one
 nonzero $\vec N^{(0)}$ normal to $\vec b^{(0)}$ since then
$\vec N^{(0)}$ times any integer is also normal to $\vec b^{(0)}$.
 The divergence in the central term is analogous
to defining the Virasoro algebra for strings in an infinite dimensional
space-time.

To deal with the divergence one can renormalize the coupling constant
$\kappa$ in ${\bf L}_{\vec N}$ by making the replacement:
\Eq{\kappa \rightarrow \kappa_R=\kappa \sqrt{{\cal D}} \;.}
If we assume a finite limit for $\kappa_R$ when
 ${\cal D}$ tends to infinity, then all that remains of
the algebra for the ${\bf L}_{\vec N}$'s is just the central term:
\Eq{[{\bf L}_{\vec N},{\bf L}_{\vec M}] \rightarrow
\frac{1}{{\cal D}} \gamma(\vec N) \;\delta_{\vec N+\vec M,(0,0,0)}  \;. }
Hence we are left with the trivial result that the renormalized
diffeomorphism generators satisfy a harmonic oscillator type algebra.

\subsection{Case A) $-$ Nonzero Windings}

Now we take up case A) when the components of $F|_{\partial D}$
(with respect to the $\theta_i$ coordinates) are not all equal to
 constants.
For simplicity we assume that
only one nontrivial mode$-$
 $$\chi^{(\vec N^{(0)})}(\vec \theta)\quad {\rm
(and\; its\; complex\; conjugate)}$$
$-$contributes to the expansion (\ref{Fexp})
for $F|_{\partial D }$.  (The complex conjugate is needed for
$F|_{\partial D }$ to be real.)
Non-zero windings means that $ \vec N^{(0)} \ne (0,0,0)$.
We now replace (\ref{V0}) by
\Eq{ \vec b^{(\vec N)} = (\delta_{\vec N,\vec N^{(0)}}
+ \delta_{\vec N,-\vec N^{(0)}})  \vec b^{(0)} \; ,
\quad \vec N^{(0)} \cdot  \vec b^{(0)} =0\;.}
For this choice we have that
$F|_{\partial D}$ is both real and closed.
 From (\ref{qNqM}) the only nonzero Poisson brackets are
\Eq{\{q_{\vec N}\;,\; q_{\vec N\pm\vec N^{(0)}}^*\}= -
(4\pi)^3i\kappa \;(2\vec N\pm\vec N^{(0)}) \cdot \vec b^{(0)} \;.
\label{qNqNb}}
Here we find that $q_{ k\vec N^{(0)}},\; kN^{(0)}_i$ being
integers for all $i$, are central charges.

  Additional central charges
may also be present in the system and this depends on the values of
 $\vec b^{(0)}$ and $\vec N^{(0)}$.  For example let us take
\Eq{\vec b^{(0)}=(\alpha,-\alpha,0)\;,\quad\vec N^{(0)}=(1,1,0) \;.}
Now if we denote $$\hat q^{N_2,N_3}_n =q_{n+N_2,N_2,N_3}\;,$$
 then all of
$\hat q^{N_2,N_3}_0$ are central charges.  The Poisson brackets for
$\hat q^{N_2,N_3}_n$, $n \ne 0$, are given by
\Eq{\{\hat q^{N_2,N_3}_n,\hat q^{M_2,M_3}_m \}
=-2(4\pi)^3i\kappa \alpha\;n \delta_{n,-m}{\cal S}_{N_2 M_2}
\delta_{N_3,-M_3}\;,   \label{hatqq}}
 where \Eq{{\cal S}_{NM}=\delta_{N+M,1}+ \delta_{M+N,-1}\;. }
This algebra is similar to what was obtained for the quantum
Hall effect.\cite{Bal}  Only here we have
 an infinite number of Landau levels labeled by $N_2$
and $N_3$.
The result is that there are an infinite number of edge currents.
Following \cite{Bal}
${\cal S}_{MN}$ can be diagonalized by a real
orthogonal matrix ${\cal M}$ (as ${\cal S}_{MN}$ is real symmetric)
and it has real eigenvalues $\lambda_{\rho}$.   Setting
\Eq{\tilde q_n^{N_3}(\rho)={\cal M}_{\rho N_2}\hat q_n^{N_2,N_3} }
we have
\Eq{\{\tilde q^{N_2}_n(\rho),\tilde q^{M_2}_m(\sigma) \}
=-2(4\pi)^3i\kappa \alpha\;n \lambda_{\rho}
 \delta_{n,-m}\delta_{\rho,\sigma}\delta_{N_3,-M_3}  \;.\label{qtil} }
(\ref{qtil}) can be readily quantized as in the previous example.

The quantum diffeomorphism generators are again normal ordered versions
of their classical counterparts (\ref{csc}).  Since as in the previous
discussion we have an infinite number of central charges (eg.,
 $q_{ k\vec N^{(0)}}$ ) we again
expect that a divergence will appear in the central term of the
diffeomorphism algebra, which upon renormalization will lead to a trivial
algebra.

\subsection{Case B)}

Finally we take up case B).  Here the only nonzero Poisson
brackets are between $q_{n,\ell,m}$ and
 $q_{-n,\ell,-m}=(-1)^m q_{n,\ell,m}^*$.
[The complex conjugation property for the $q$'s follows from the
complex conjugation property for the spherical harmonics,
$Y^{m*}_{\ell}(\theta,\phi)= (-1)^m Y^{-m}_{\ell}(\theta,\phi)$.]
They are given by
\Eq{\{q_{n,\ell,m}\;,\; q^*_{n,\ell,m}\}= -16\pi ig\kappa \;  n\; \;.}
The corresponding quantum mechanical commutation relations are
\Eq{[Q_{n,\ell,m}\;,\; Q^{\dagger}_{n,\ell,m}]=
 16\pi g\kappa \;  n\; \;,}
$Q_{n,\ell,m}$ being the operators corresponding to $q_{n,\ell,m}$.
Now if $ g\kappa >0  \;\;(g \kappa <0 ),\;Q_{n,\ell,m}$ for
$n \;>0 \;(n \;<0)$ are annihilation operators (upto a normalization) and
$Q_{n,\ell,m}^\dagger $ creation operators.
{\it We note that there is no quantization of the magnetic charge $g$ for
this system.}

The quantum diffeomorphism generators are now normal ordered versions
of (\ref{SugB}).  As in the previous examples there are
an infinite number of central charges ($Q_{0,\ell,m}$) and
we thus expect that a divergence will appear in the central term of the
diffeomorphism algebra, leading to a trivial algebra upon
 renormalization.

\section{Concluding Remarks}

Here we remark on possible generalizations of our work.

The first natural generalization is to go to the case of
the nonabelian Chern-Simons theory in five dimensions.
The constraint analysis for this case should proceed in an analogous
fashion to that in Sec 2.1.  A complication however arises in the
imposition of boundary conditions on the fields.  In the abelian
theory we needed to require that the field strengths are fixed at
$\partial D$ in order that the observables $q(f)$ be differentiable.
 In the nonabelian case however such a boundary condition
has no gauge invariant meaning.  It appears that there we must instead
fix an entire orbit in the space of field strengths.  Such orbits are
generated by gauge transformations at $\partial D$ and correspond to
the classical degrees of freedom of the edge states.  An effective
field theory for these degrees of freedom should be a four dimensional
analogue of the Wess-Zumino-Witten model.
Furthermore the algebra of observables which one finds
there should be analogous to the nonabelian Kac-Moody algebra.

Another possible extension of this paper involves admitting
additional topological actions in five dimensions.  For this we can
introduce a {\it set} of potentials one-forms $\{A^{(1)},A^{(2)},$
$A^{(3)},...A^{(n)}\}$ on the five-manifold $M$.
We can then generalize the action (\ref{tcsa}) to
\Eq{\kappa_{abc}\int_{M} A^{(a)}\wedge dA^{(b)}\wedge dA^{(c)} \;,}
where $\kappa_{abc}$ is symmetric with respect to all three indices.
This is analogous to the Chern-Simons description of the quantum
Hall effect where there
$a,b$ and $c$ label the $a^{th},b^{th}$ and $c^{th}$
Landau level.\cite{Bal}

Rather than introducing more
one forms on $M$ we can define $p(>1)-$forms: $B^{(p)}$.
The following topological actions are then possible:
\Eq{\int_M A\wedge B^{(2)} \wedge B^{(2)}\;, \label{mtaf}}
\Eq{\int_M A\wedge F \wedge B^{(2)}       \;,}
\Eq{\int_M A\wedge d B^{(3)}               \;,  }
\Eq{\int_M B^{(2)}\wedge  B^{(3)}         \;,   }
\Eq{\int_M A\wedge  B^{(4)} \;.\label{mtal}}
In general the action may be a linear combination of (\ref{tcsa}) and
(\ref{mtaf}-\ref{mtal}).   With such a modification to the original
dynamics there exists the possibility that the boundary conditions
(\ref{bcf}) on the field strengths
 may be relaxed and that further we may obtain
more than just a single scalar degree of freedom on $\partial D$.

In a future work we plan to address some of the above modifications
to the five dimensional Chern-Simons system.

\bigskip
{\bf Acknowledgement}
A. S. was supported during the course of this work
by the Department of Energy, USA, under contract number
DE-FG05-84ER40141. K. G. was supported in part by the
U.S. Department of Energy under Grant No.
DE-FG02-87ER40371.  K. G. also acknowledges the support of
the Department of Physics of the University of Rochester
where a part of this work was conducted.
We wish to thank A. P. Balachandran, V. P. Nair and
S. Sarker for very helpful discussions.

\end{document}